\documentclass{PoS}

\newcommand{\cO}{{\cal O}}
\newcommand{\as}{\alpha_s}
\newcommand{\wh}{\widehat}
\newcommand{\nn}{\nonumber}
\newcommand{\eqn}[1]{(\ref{#1})}
\newcommand{\gev}{\mbox{\rm GeV}}
\newcommand{\tvs}{\vbox{\vskip 4mm}}

\title{Determination of $\alpha_s$ from $\tau$ decays}

\ShortTitle{$\as$ from $\tau$ decays}

\author{\speaker{Matthias Jamin}\\
        Instituci\'o Catalana de Recerca i Estudis Avan\c{c}ats (ICREA), IFAE,\\
        Universitat Auton\`oma de Barcelona, E-08193 Bellaterra,
        Barcelona, Spain\\
        E-mail: \email{jamin@ifae.es}}

\abstract{Hadronic $\tau$ decays offer the possibility of determining the
strong coupling $\as$ at relatively low energy. Precisely for this reason,
however, good control over the perturbative QCD corrections, the
non-perturbative condensate contributions in the framework of the operator
product expansion (OPE), as well as the corrections going beyond the OPE, the
duality violations (DVs), is required. On the perturbative QCD side, the
contour-improved versus fixed-order resummation of the series is still an
issue, and will be discussed. Regarding the analysis, self-consistent fits
to the data including all theory parameters have to be performed, and this
is also explained in some detail. The fit quantities are moment integrals
of the $\tau$ spectral function data in a certain energy window and care
should be taken to have acceptable perturbative behaviour of those moments
as well as control over higher-dimensional operator corrections in the OPE.}

\FullConference{Xth Quark Confinement and the Hadron Spectrum,\\
		October 8-12, 2012\\
		TUM Campus Garching, Munich, Germany}

\begin{document}

\section{Introduction}

The $\tau$ lepton is the only known lepton heavy enough to also decay
hadronically. In view of its mass of $M_\tau\approx 1.8\,\gev$, it provides
an excellent laboratory for the investigation of low-energy QCD. On the other
hand, the scale is so low that the inclusion of effects beyond perturbative
QCD is required for precision studies. The non-perturbative effects appear in
the framework of the operator product expansion (OPE) as vacuum-condensate
terms, or contributions beyond the OPE, named {\em duality violations}, since
they signal the breakdown of the quark-hadron duality picture of the OPE.

The central experimental observable is the total $\tau$ hadronic width
\begin{equation}
\label{Rtauex}
R_\tau \,\equiv\, \frac{\Gamma[\tau^- \to {\rm hadrons} \, \nu_\tau (\gamma)]}
{\Gamma[\tau^- \to e^- \overline \nu_e \nu_\tau (\gamma)]} \,=\,
3.6280(94) \;\cite{hfag12} \,,
\end{equation}
which was explored in the seminal theoretical analyses \cite{bnp92,bra88,np88}
as a means to determine the strong coupling $\as$. In subsequent years, the
acquisition of large data sets by the ALEPH and OPAL collaborations allowed
the extraction of inclusive spectral $\tau$-decay distributions, the so-called
spectral functions, and their separation into the light-quark (up and down)
vector and axialvector channels, as well as the strange channel
\cite{aleph98,opal98,aleph05}. The additional experimental information can
be employed to compute moments of the decay spectra which yield further data
points to be compared to theory. In principle, also exclusive decay
distributions are available and could be investigated, though they are less
useful for basic QCD studies.

On the theoretical side, recent years have seen several improvements in
the description of hadronic $\tau$ decays. The most important one was the
impressive analytical computation of the perturbative order $\alpha_s^4$
correction \cite{bck08}, which has revived the interest in $\alpha_s$ analyses
from $\tau$ decays. Generally, one might think that an additional order in
the perturbative expansion would reduce the theoretical uncertainties. However,
the theoretical expectation for the total hadronic width and moments of the
decay spectra depends on the way that large logarithms appearing in the
expansion are resummed by means of the renormalisation group (RG). The two
most commonly used approaches are fixed-order perturbation theory (FOPT), and
contour-improved perturbation theory (CIPT) \cite{piv91,dp92}, and the newly
available $\cO(\alpha_s^4)$ correction has made the dispersion in the
theoretical prediction only more distinct. Hence, efforts in recent years have
been devoted to better understand the origin of the differences in the RG
improvement and attempts to single out the more reliable procedure
\cite{jam05,bj08,cf09,dm10,aacf12,bbj12}. Those shall be reviewed in more
detail below.

The second main step forward has been a proper inclusion of violations of
quark-hadron duality in the $\tau$ sum rule analysis. The calculation of the
theoretical side of the sum rule involves an integral over a circle in the
complex energy plane with radius $s_0=M_\tau^2$ or smaller (down to about
$1.5\,\gev^2$). In the region close to the physical axis with real, positive
$s_0$, the OPE breaks down and contributions beyond it become relevant. This
should be clear, as the OPE is unable to directly describe the hadronic
resonance structure. Since a sound theoretical description of duality violations
(DVs) is not available, a initial model was laid out in ref.~\cite{bsz01}, and
later refined in refs.~\cite{cgp05,cgp08}. A complete analysis requires the
simultaneous, self-consistent determination of all occurring parameters, that
is, $\as$, the OPE condensate parameters, as well as the DV model parameters.
On the basis of the OPAL data, such an analysis was performed in
refs.~\cite{boito11,boito12}, the discussion of which will be the second main
topic of this writeup.

\section{The anatomy of $R_{\tau,V+A}$}

Most suitable for a determination of $\as$ are the $\tau$ decay rates into
light $u$ and $d$ quarks $R_{\tau,V/A}$ via the vector or axialvector current,
and the related moments, since in this case power corrections are especially
suppressed. Theoretically, $R_{\tau,V/A}$ can be expressed as
\begin{equation}
\label{RtauVA}
R_{\tau,V/A} \,= \frac{N_c}{2}\,|V_{ud}|^2\,S_{\rm EW}\,\Big[\,
1 + \delta^{(0)} + \sum\limits_{D\geq 2} \delta_{ud,V/A}^{(D)} +
\delta_{V/A}^{\rm DV} \,\Big]\,,
\end{equation}
where $S_{\rm EW}=1.0201(3)$ \cite{ms88,bl90,erl02} comprises electroweak
corrections, $\delta^{(0)}$ denotes the perturbative QCD correction, the
$\delta_{ud,V/A}^{(D)}$ are quark mass and higher $D$-dimensional operator
corrections which arise in the framework of the OPE, and $\delta_{V/A}^{\rm DV}$
is the DV contribution beyond the OPE. Before entering a more detailed
discussion of the $\as$ analysis, let us review our current knowledge regarding
the anatomy of $R_{\tau,V+A}$, which takes the general form
\begin{equation}
\label{RtauVpA}
R_{\tau,V+A} \,= 3\,|V_{ud}|^2\,S_{\rm EW}\,\Big[\,
1 + \delta^{(0)} + \delta_{V+A}^{\rm NP} \,\Big]\,.
\end{equation}
The higher-dimensional OPE and the DV corrections have been lumped together
into the non-perturbative correction $\delta_{V+A}^{\rm NP}$. To obtain the
general picture, first we assume $\as$ to be known. Evolving the present
PDG average $\as(M_Z)=0.1184(7)$ to the $\tau$ mass, yields
$\as(M_\tau)=0.3186(58)$. Even though the PDG average also includes $\as$
from $\tau$'s, removing this datum only has a small influence, so that it
appears justified to employ the PDG value.

To derive our expectation for $\delta^{(0)}$, which gives the dominant
correction to $R_{\tau,V+A}$, the theoretical expressions in FOPT and CIPT are
briefly reminded. In FOPT the fixed renormalisation scale $\mu=M_\tau$ is
chosen, which results in \cite{jam05}
\begin{equation}
\label{del0FO}
\delta^{(0)}_{\rm FO} \,=\, \sum\limits_{n=1}^\infty a(M_\tau^2)^n
\sum\limits_{k=1}^{n} k\,c_{n,k}\,J_{k-1} \,,
\end{equation}
where $a(\mu^2)\equiv\as(\mu)/\pi$, and $c_{n,k}$ are coefficients which
appear in the perturbative expansion of the vector correlation function.
At each perturbative order, the coefficients $c_{n,1}$ can be considered
independent, while all other $c_{n,k}$ with $k\geq 2$ are calculable from
the RG equation. Explicit results for the analytically known $c_{n,1}$ can be
found in ref.~\cite{bck08}. Furthermore, the $J_l$ are contour integrals in
the complex $s$-plane which for example are presented in \cite{jam05}.

In CIPT, on the other hand, an $s$-dependent renormalisation scale is
introduced, which partially resums higher-order effects, namely the ones
due to the running of $\as$, leading to \cite{piv91,dp92}
\begin{equation}
\label{del0CI}
\delta^{(0)}_{\rm CI} \,=\, \sum\limits_{n=1}^\infty c_{n,1}\,
J_n^a(M_\tau^2)
\end{equation}
in terms of the contour integrals $J_n^a(M_\tau^2)$ over the running coupling,
defined as:
\begin{equation}
\label{Jna}
J_n^a(M_\tau^2) \,\equiv\, \frac{1}{2\pi i} \!\!\oint\limits_{|x|=1}\!\!
\frac{dx}{x}\,(1-x)^3\,(1+x)\,a^n(-M_\tau^2 x) \,.
\end{equation}
In contrast to FOPT, for CIPT each order $n$ only depends on the corresponding
coefficient $c_{n,1}$. All contributions proportional to the coefficient
$c_{n,1}$, which in FOPT appear at all perturbative orders equal or greater
$n$, are resummed into a single term.

Numerically, the two approaches lead to significant differences. Employing
the value for $\as(M_\tau)$ given above, one obtains
\begin{eqnarray}
\label{del0FOn}
\delta^{(0)}_{\rm FO} &\!=\!& 0.2022 \pm 0.0069 \pm 0.0030
\,=\, 0.2022(75) \,, \\
\tvs
\label{del0CIn}
\delta^{(0)}_{\rm CI} &\!=\!& 0.1847 \pm 0.0048 \pm 0.0033
\,=\, 0.1847(58) \,,
\end{eqnarray}
where the first error corresponds to the uncertainty in $\as(M_\tau)$ and
the second to an estimate of higher orders through a variation of the
coefficient $c_{5,1} = 283 \pm 283$ \cite{bj08}. Given the results for
$\delta^{(0)}$, the value $|V_{ud}|=0.97425(22)$ \cite{th08}, as well as
$R_{\tau,V+A}=3.4671(82)$, which follows from $B_{VA}=61.85(11)\%$ and
$B_e^{\rm uni}=17.839(28)\%$ \cite{hfag12}, one can estimate
$\delta_{V+A}^{\rm NP}$, with the finding
\begin{equation}
\delta_{V+A,{\rm FO}}^{\rm NP} \,=\, -\,0.0086(80) \,, \qquad
\delta_{V+A,{\rm CI}}^{\rm NP} \,=\,    0.0089(65) \,.
\end{equation}
This analysis shows that $\delta_{V+A}^{\rm NP}$ is expected to be
$\lesssim 1\%$ and from this simple estimate compatible with zero at about
$1\sigma$. Hence, the non-perturbative correction is much smaller than the
perturbative one. Still, at the current level of precession, where the error on
the PDG average for $\as$ induces a shift in $\delta^{(0)}$ of roughly $0.5\%$,
the non-perturbative contribution becomes relevant. Before real progress in the
determination of $\as$ from $\tau$ decays can be made, however, the difference
between FOPT and CIPT has to be settled first, as it is certainly more
important.

\section{Adler function at higher orders}

The question whether FOPT or CIPT provides a better approximation to
$\delta^{(0)}$ hinges on the behaviour of the vector correlator, or
equivalently the Adler function, at higher orders. To make progress in this
direction, additional information beyond the analytically known orders has
to be taken into account. An attempt towards this goal, based on a model for
the Borel-transformed Adler function, was presented in ref.~\cite{bj08} and
shall be described next.

The perturbative expansion of the Adler function $D(s)$ takes the form
\begin{equation}
\label{DVs}
4\pi^2\,D(s) \,\equiv\, 1 + \wh D(s) \,=\, \sum\limits_{n=0}^\infty
c_{n,1}\,a(s)^n \,.
\end{equation}
For the following it is slightly more convenient to utilise the function
$\wh D(s)$ instead of $D(s)$. Its Borel transform $B[\wh D](t)$ is defined by
\begin{equation}
\label{Dalpha}
\wh D(\alpha) \,\equiv\, \int\limits_0^\infty dt\,{\rm e}^{-t/\alpha}\,
B[\wh D](t)\,.
\end{equation}
The integral $\wh D(\alpha)$, if it exists, gives the Borel sum of the original
divergent series. It was found that the Borel-transformed Adler function
$B[\wh D](t)$ obtains infrared (IR) and ultraviolet (UV) renormalon poles at
positive and negative integer values of the variable $u\equiv 9t/(4\pi)$,
respectively \cite{ben93,bro93,ben98}. (With the exception of $u=1$, since
there is no corresponding gauge-invariant $D=2$ operator.)

Guided by the large-$\beta_0$ approximation \cite{ben98}, the influence of
renormalon poles on the perturbative expansion should be as follows:
asymptotically, that is for high orders, the perturbative expansion is
dominated by the $u=-1$ UV pole that is closest to $u=0$. At intermediate
orders some dominance of the low-lying IR poles ($u=2$, $u=3$) is observed,
while very low orders are no longer dominated by a small subset of renormalon
poles. This suggests the following ansatz \cite{bj08}
\begin{equation}
\label{BRu}
B[\wh D](u) \,=\, B[\wh D_1^{\rm UV}](u) + B[\wh D_2^{\rm IR}](u) +
B[\wh D_3^{\rm IR}](u) + d_0^{\rm PO} + d_1^{\rm PO} u \,,
\end{equation}
which includes one UV renormalon at $u=-1$, the two leading IR renormalons at
$u=2$ and $u=3$, as well as polynomial terms for the two lowest perturbative
orders. Explicit expressions for the UV and IR renormalon pole terms
$B[\wh D_p^{\rm UV}](u)$ and $B[\wh D_p^{\rm IR}](u)$ can be found in
section~5 of ref.~\cite{bj08}.

Apart from the residues $d_p^{\rm UV}$ and $d_p^{\rm IR}$, the full structure
of the renormalon pole terms is dictated by the OPE and the RG. Therefore, the
model~\eqn{BRu} depends on five parameters, the three residua $d_1^{\rm UV}$,
$d_2^{\rm IR}$ and $d_3^{\rm IR}$, as well as the two polynomial parameters
$d_0^{\rm PO}$ and $d_1^{\rm PO}$. These parameters can be fixed by matching
to the perturbative expansion of $\wh D(s)$ up to $\cO(\as^5)$, whereby also
the estimate for $c_{5,1}=283$ is required. The parameters of the model
\eqn{BRu} then are found to be:
\begin{equation}
\label{dUVIRa}
d_1^{\rm UV} =\, -\,1.56\cdot 10^{-2} \,, \qquad
d_2^{\rm IR} =\,    3.16  \,, \qquad
d_3^{\rm IR} =\, -\,13.5 \,, \nn \\[-1mm]
\end{equation}
\begin{displaymath}
\label{dUVIRb}
d_0^{\rm PO} =\,    0.781 \,, \qquad
d_1^{\rm PO} =\,    7.66\cdot 10^{-3} \,. \nn
\end{displaymath}
The fact that the parameter $d_1^{\rm PO}$ turns out small implies that the
coefficient $c_{2,1}$ is already reasonably well described by the renormalon
pole contribution, although it was not used to fix the residua.

\begin{figure}[htb]
\begin{center}
\includegraphics[angle=0, width=11cm]{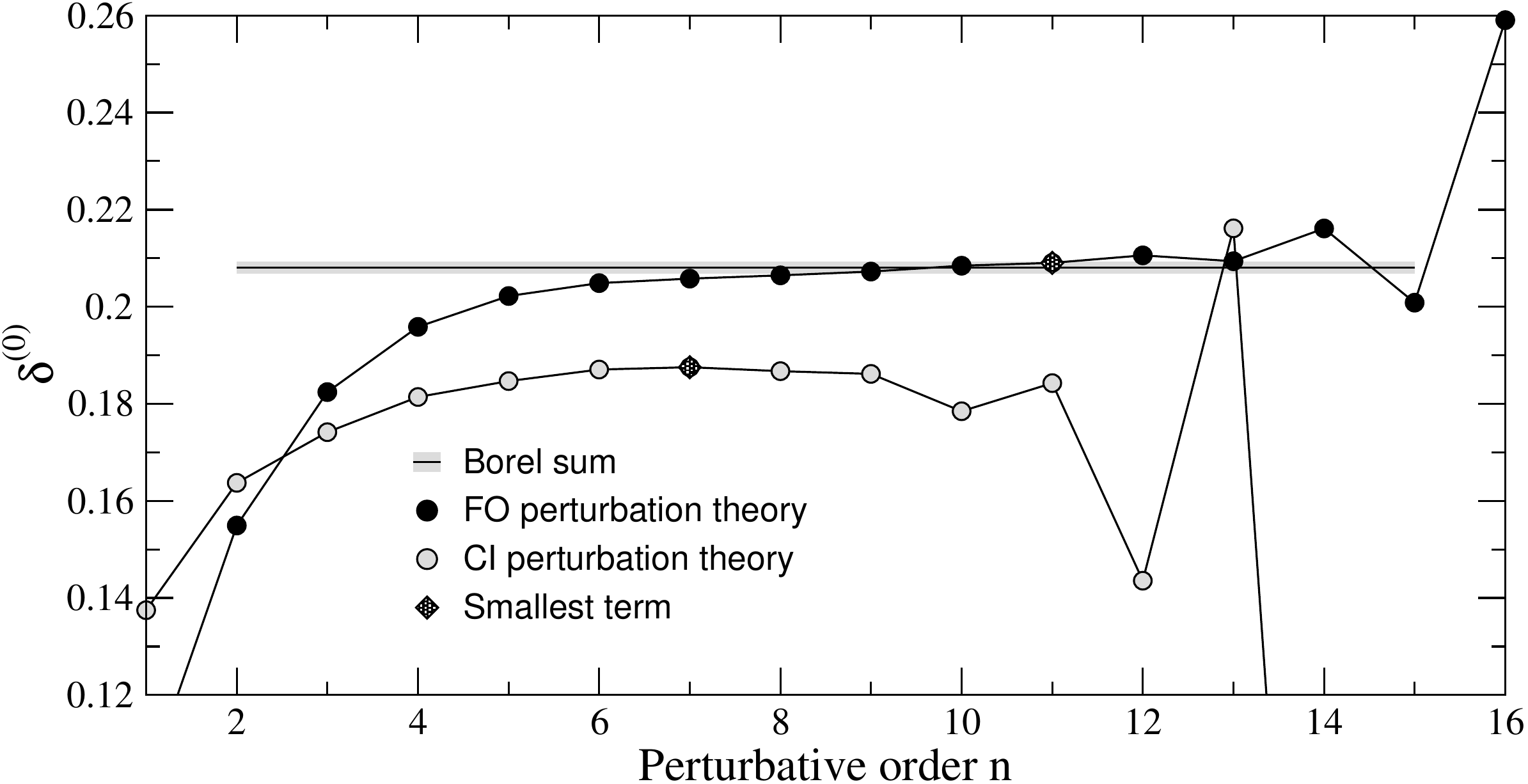}
\caption{Results for $\delta^{(0)}_{\rm FO}$ (full circles) and
$\delta^{(0)}_{\rm CI}$ (grey circles) at $\as(M_\tau)=0.3186$, employing the
model (3.3), as a function of the order $n$ up to which the terms in the
perturbative series have been summed. The straight line represents the result
for the Borel sum of the series.}
\vspace{-2mm}
\end{center}
\end{figure}

The implications of the model \eqn{BRu} for $\delta^{(0)}$ in FOPT and CIPT
is graphically represented in figure~1. The full circles denote the
result for $\delta^{(0)}_{\rm FO}$ and the grey circles the one for
$\delta^{(0)}_{\rm CI}$, as a function of the order $n$ up to which the
perturbative series has been summed. The straight line corresponds to the
principal value Borel sum of the series. The order at which the series have
their smallest terms is indicated by the grey diamonds. As is evident from
figure~1, FOPT displays the behaviour expected from an asymptotic series: the
terms decrease up to a certain order around which the closest approach to the
resummed result is found, and for even higher orders, the divergent large-order
behaviour of the series sets in. For CIPT, on the other hand, the asymptotic
behaviour sets in earlier, and the series is never able to come close to the
Borel sum. In the large-$\beta_0$ approximation, this was already observed in
ref.~\cite{bbb95}.

The superiority of FOPT over CIPT critically depends on the size of the $u=2$
residue $d_2^{\rm IR}$. If this residue in real QCD would turn out substantially
smaller than in eq.~\eqn{dUVIRa}, CIPT could provide the better approach to the
resummed series. Corresponding models have been studied in ref.~\cite{bbj12}
whose main aim was to investigate the perturbative behaviour of a large set of
moments used in $\as$ analyses from $\tau$ decays within the Borel models.
Nevertheless, there is no known mechanism in QCD to suppress the $u=2$ residue
and thus the behaviour that favours FOPT appears more likely. Other studies
that implement information on the renormalon structure of QCD include
refs.~\cite{cf09,aacf12}, where also conformal mappings were applied to the
series.

\section{Self-consistent fits, OPE and duality violations}

A self-consistent analysis of $\as$ from $\tau$ decays includes the simultaneous
determination of all other parameters entering the game. These include QCD
condensates as well as DV parameters. The required additional information
can be obtained through the use of weighted integrals of the inclusive decay
spectra up to an energy $s_0\leq M_\tau^2$, the so-called moments. Making
use of the fact that the corresponding correlation functions are analytic in
the complex $s$-plane except for a cut along the real axis, one can define
the moments via \cite{boito11,boito12}
\begin{equation}
\label{Rtaucon}
R^{w_i}_{V/A}(s_0) \,=\, 6\pi i\,S_{\rm EW} |V_{ud}|^2 \!\!\!
\oint\limits_{|s|=s_0} \frac{d s}{s_0}\, w_i\left(s\right) \biggl[\,\Pi^{(1+0)}_{V/A}(s) +
\frac{2s}{(s_0+2s)}\,\Pi^{(0)}_{V/A}(s) \,\biggr] \,,
\end{equation}
where $\Pi^{(1)}_{V/A}$ and $\Pi^{(0)}_{V/A}$ are spin-1 and spin-0 mesonic
correlators, and the particular case of eq.~\eqn{RtauVA} corresponds to
$R^{w_\tau}_{V/A}(M_\tau^2)$ with the kinematic weight
$w_\tau(s) = (1-s/M_\tau^2)^2(1+2s/M_\tau^2)$.

Typically, in the past, moment analyses of $\tau$ decay spectra were based
on 5 moments to determine 4 parameters, $\as$, the gluon condensate
$\langle\as GG\,\rangle$, as well as the $D=6$ and $D=8$ OPE corrections
\cite{aleph98,opal98,aleph05,ddhmz08}. These analyses, however, suffer from
several deficiencies: the used moments are all calculated at $s_0=M_\tau^2$.
In ref.~\cite{my08} it was then shown that the fit solutions are not stable
under a variation of $s_0$ towards lower values. Next, because pinched-weights
with zeros at $s=s_0$ were used, it was assumed that DVs are negligible. Since
the DVs enter differently in the different moments, this is potentially
dangerous. And finally, the perturbative behaviour of some of the used moments
is very bad \cite{bbj12}, whence they should be avoided in the $\as$ analysis.
Besides, the ALEPH analyses of refs.~\cite{aleph05,ddhmz08} were based on data
with an incomplete covariance matrix \cite{bcgjmop10}, and it remains open
what is the impact of this shortcoming.

The aim of refs.~\cite{boito11,boito12} was to improve on all the above
deficiencies. Firstly, only OPAL data \cite{opal98} were used, employing the
original set in ref.~\cite{boito11} and an updated set incorporating present
day $\tau$ branching fractions in ref.~\cite{boito12}. Next, DVs were included
in the fit which required the use of one un-pinched moment particularly
sensitive to DVs, the simplest choice being $w(s)=1$. In multi-moment fits
the additional moments were required to have good perturbative behaviour,
which excluded moments sensitive to the gluon condensate, and to suppress $D>8$
OPE terms. And finally, the $s_0$-dependence of the moments was included down
to about $s_0\approx 1.5\,\gev^2$. Figure~2 displays the outcome of the most
basic fit with the single moment $w(s)=1$ and just to the vector-channel
spectrum. Also with regard to the OPE, this fit is cleanest as all OPE
corrections are additionally suppressed. In ref.~\cite{boito12} the resulting
$\as$ was found to be:
\begin{equation}
\label{astau}
\as(M_\tau)=0.325(18) \;\;{\rm (FOPT)} \,, \qquad
\as(M_\tau)=0.347(25) \;\;{\rm (CIPT)} \,.
\end{equation}
The uncertainties in \eqn{astau} turn out to be substantially larger than in
previous $\as$ analyses from the $\tau$. One reason is the use of OPAL data,
but more importantly employing $w(s)=1$. Naively one might think that the
inclusion of further moments into the fit would reduce the error on $\as$.
As demonstrated in refs.~\cite{boito11,boito12}, however, this is not so,
because all moments are strongly correlated. Even combined fits to vector
and axialvector with up to three weight functions only resulted in a mild
reduction in the uncertainty on $\as$.

\begin{figure}[!t]
\begin{center}
\includegraphics[width=7cm]{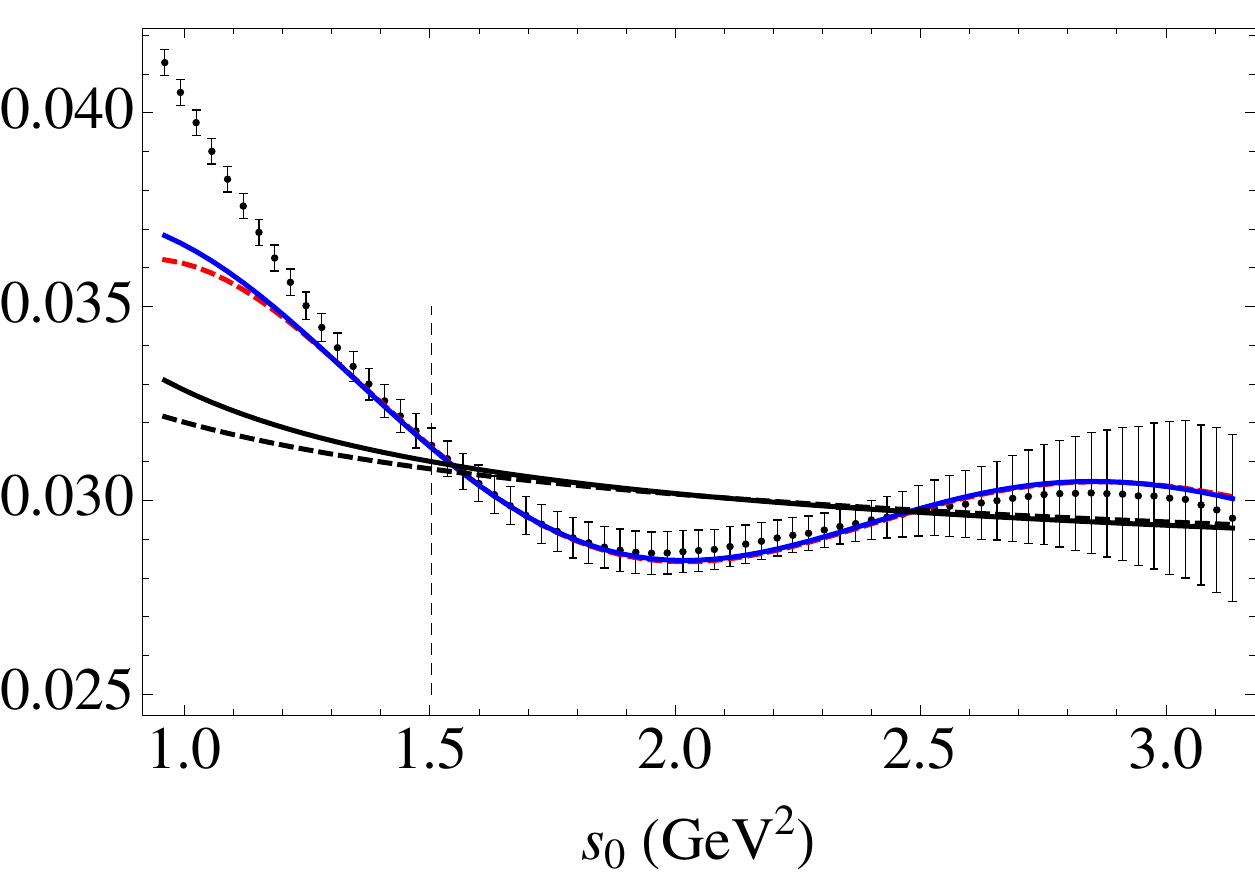}
\hspace{2mm}
\includegraphics[width=7cm]{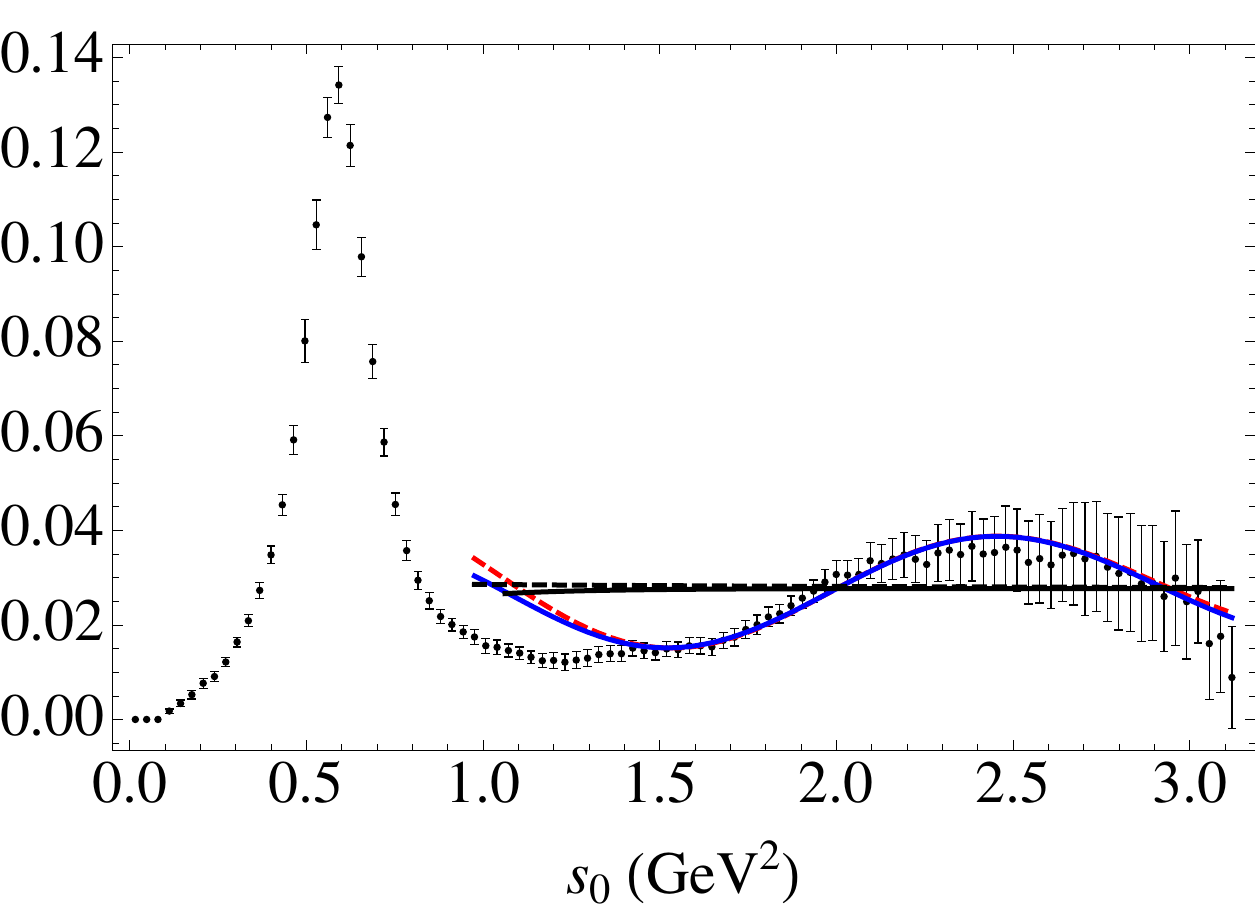}
\caption{\it Vector channel fit to the moments $R_V^{w=1}(s_0)$. Left panel:
comparison of the experimental moments with the theoretical fit curves for FOPT
in blue (solid) and CIPT in red (dashed). The (much flatter) black curves show
the pure OPE parts without DVs. Right panel: comparison of the fit prediction
for the vector spectral function with the experimental OPAL data.}
\vspace{-2mm}
\end{center}
\end{figure}

In conclusion, it appears as if only better $\tau$ spectral function data,
either from a revised ALEPH analysis, or from the $B$-factories BaBar and
Belle, will help to improve the situation, together with a resolution of
the CIPT versus FOPT controversy of treating higher perturbative orders.

\vskip 4mm 
\section*{Acknowledgements} \vspace{-2mm}
MJ has been supported in parts by the Spanish Ministry (grants
CICYT-FEDER FPA2011-25948, CPAN CSD2007-00042) and by the Catalan Government
(grant SGR2009-00894).


\begin{thebibliography}{99}

\bibitem{hfag12}
  Y.~Amhis {\it et al.} [Heavy Flavour Averaging Group],
  arXiv:1207.1158 [hep-ex].

\bibitem{bnp92}
E.~Braaten, S.~Narison, and A.~Pich, {\em Nucl. Phys.} {\bf B373} (1992) 581.

\bibitem{bra88}
E.~Braaten {\em Phys. Rev. Lett.} {\bf 60} (1988) 1606;
           {\em Phys. Rev.} {\bf D39} (1989) 1458.

\bibitem{np88}
S.~Narison and A.~Pich, {\em Phys. Lett.} {\bf B211} (1988) 183.

\bibitem{aleph98}
{\bf ALEPH} Collaboration, R.~Barate {\em et~al.}, {\em Eur. Phys. J.}
{\bf C4} (1998) 409.

\bibitem{opal98}
{\bf OPAL} Collaboration, K.~Ackerstaff {\em et~al.}, {\em Eur. Phys. J.}
{\bf C7} (1999) 571 [hep-ex/9808019].

\bibitem{aleph05}
{\bf ALEPH} Collaboration, S.~Schael {\em et~al.}, {\em Phys. Rept.} {\bf 421}
(2005) 191 [hep-ex/0506072].

\bibitem{bck08}
P.A. Baikov, K.G. Chetyrkin, and J.H. K{\"u}hn, {\em Phys. Rev. Lett.}
{\bf 101} (2008) 012002, arXiv:0801.1821 [hep-ph].

\bibitem{piv91}
A.A. Pivovarov, {\em Z. Phys.} {\bf C53} (1992) 461 [hep-ph/0302003].

\bibitem{dp92}
F.~{Le Diberder} and A.~Pich, {\em Phys. Lett.} {\bf B286} (1992) 147.

\bibitem{jam05}
M.~Jamin, {\em JHEP} {\bf 0509} (2005) 058 [hep-ph/0509001].

\bibitem{bj08}
M.~Beneke and M.~Jamin, {\em JHEP} {\bf 09} (2008) 044,
arXiv:0801.1821 [hep-ph].

\bibitem{cf09}
I.~Caprini and J.~Fischer, {\em Eur. Phys. J.} {\bf C64} (2009) 35,
arXiv:0906.5211 [hep-ph]; {\em Phys. Rev.} {\bf D84} (2011) 054019,
arXiv:1106.5336 [hep-ph].

\bibitem{dm10}
S.~Descotes-Genon and B.~Malaescu, arXiv:1002.2968 [hep-ph].

\bibitem{aacf12}
G.~Abbas, B.~Ananthanarayan, I.~Caprini, and J.~Fischer,
{\em Phys. Rev.} {\bf D87} (2013) 014008, arXiv:1211.4316 [hep-ph].

\bibitem{bbj12}
M.~Beneke, D.~Boito and M.~Jamin, {\em JHEP} {\bf 1301} (2013) 125,
arXiv:1210.8038 [hep-ph]; D.~Boito, these proceedings, arXiv:1301.3008 [hep-ph].

\bibitem{bsz01}
B.~Blok, M.A.~Shifman and D.X.~Zhang, {\em Phys. Rev.} {\bf D57} (1998) 2691
[hep-ph/9709333], Erratum-ibid. {\bf D59} (1999) 019901.

\bibitem{cgp05}
O.~Cat\`a, M.~Golterman, and S.~Peris, {\em JHEP} {\bf 0508} (2005) 076
[hep-ph/0506004].

\bibitem{cgp08}
O.~Cat\`a, M.~Golterman, and S.~Peris, {\em Phys. Rev.} {\bf D77} (2008) 093006,
arXiv:0803.0246 [hep-ph]; {\em Phys. Rev.} {\bf D79} (2009) 053002,
arXiv:0812.2285 [hep-ph].

\bibitem{boito11}
D.~Boito, O.~Cat\`a, M.~Golterman, M.~Jamin, K.~Maltman, J.~Osborne,
and S.~Peris, {\em Phys. Rev.} {\bf D84} (2011) 113006,
arXiv:1110.1127 [hep-ph].

\bibitem{boito12}
D.~Boito, M.~Golterman, M.~Jamin, A.~Mahdavi, K.~Maltman, J.~Osborne,
and S.~Peris, {\em Phys. Rev.} {\bf D85} (2012) 093015,
arXiv:1203.3146 [hep-ph].

\bibitem{ms88}
W.~Marciano and A.~Sirlin, {\em Phys. Rev. Lett.} {\bf 61} (1988) 1815.

\bibitem{bl90}
E.~Braaten and C.~S. Li, {\em Phys. Rev.} {\bf D42} (1990) 3888.

\bibitem{erl02}
J.~Erler, {\em Rev. Mex. Fis.}  {\bf 50} (2004) 200 [hep-ph/0211345].

\bibitem{pdg12}
J.~Beringer et al. (Particle Data Group), {\em Phys. Rev.} {\bf D86}
(2012) 010001.

\bibitem{th08}
J.C.~Hardy and I.S.~Towner, {\em Phys. Rev.} {\bf C79} (2009) 055502,
arXiv:0812.1202 [nucl-ex].

\bibitem{ben93}
M.~Beneke, {\em Nucl. Phys.} {\bf B405} (1993) 424.

\bibitem{bro93}
D.~J. Broadhurst, {\em Z. Phys.} {\bf C58} (1993) 339.

\bibitem{ben98}
M.~Beneke, {\em Phys. Rept.} {\bf 317} (1999) 1 [hep-ph/9807443].

\bibitem{bbb95}
P.~Ball, M.~Beneke, V.M.~Braun,
{\em Nucl. Phys.} {\bf B452} (1995) 563 [hep-ph/9502300].

\bibitem{ddhmz08}
M.~Davier, S.~Descotes-Genon, A.~H{\"o}cker, B.~Malaescu, and Z.~Zhang,
{\em Eur. Phys. J.} {\bf C56} (2008) 305, arXiv:0803.0979 [hep-ph].

\bibitem{my08}
K.~Maltman and T.~Yavin, {\em Phys. Rev.} {\bf D78} (2008) 094020,
arXiv:0807.0650 [hep-ph].

\bibitem{bcgjmop10}
D.~Boito, O.~Cat\`a, M.~Golterman, M. Jamin, K. Maltman, J. Osborn, and
S. Peris, {\em Nucl. Phys. Proc. Suppl.} {\bf 218} (2011) 104,
arXiv:1011.4426 [hep-ph].

\end{thebibliography}
\end{document}